\documentclass[aps,prl]{revtex4}
\def\ee{\end{equation}}
\def\bea{\begin{eqnarray}}

\def\bra#1{\langle #1 |}
\def\ket#1{| #1\rangle}
\def\ul#1{ \underline{#1}}

\def\braopket#1#2#3{\langle \, #1 \, | \, #2 \, | \, #3 \rangle}

\begin{document}

\title{Quantum Reality via Late Time Photodetection}

\author{Adrian \surname{Kent}}
\affiliation{Centre for Quantum Information and Foundations, DAMTP, Centre for
  Mathematical Sciences, University of Cambridge, Wilberforce Road,
  Cambridge, CB3 0WA, U.K.}
\affiliation{Perimeter Institute for Theoretical Physics, 31 Caroline Street North, Waterloo, ON N2L 2Y5, Canada.}
\email{A.P.A.Kent@damtp.cam.ac.uk} 

\date{August 2016} 

\begin{abstract}
We further investigate postulates for realist versions of  
relativistic quantum theory and quantum field theory in Minkowski
space and other background space-times.   According to these postulates, quantum theory is supplemented
by local variables that depend on possible outcomes of hypothetical
measurements on the late time electromagnetic field in spacelike
separated regions.  We illustrate the implications in simple examples using photon 
wave mechanics, and discuss possible extensions to quantum field theory. 

\end{abstract}
\maketitle
  
\section{Introduction}

In previous papers \cite{kent2012real,kent2014solution,kent2015lorentzian}, we
described ideas aimed at defining realist and
Lorentz covariant versions of relativistic quantum
theory or quantum field theory in Minkowski space.
The definitions also extend to other background
spacetimes with appropriate asymptotic properties
in the far future.
These papers set out a variety of possible ways of defining 
variables that extend quantum theory and may allow a description
of physical reality consistent with our observations, including
the observation that measurements have single definite outcomes.
These variables are defined at
each point in space-time, but do not follow local 
equations of motion.   Following Bell's terminology, we refer to
them as {\it beables}.     

In the main part of this paper, we focus on the ideas proposed in
Ref. \cite{kent2015lorentzian}, according to which the beables at any point
$x$ in space-time depend on the initial quantum state and also on
the outcomes of fictitious measurements made on the
electromagnetic field on asympotically late time surfaces that are space-like separated from
$x$.   
We begin to flesh out this proposal and its implications, using somewhat more realistic models than
the toy models discussed in Ref. \cite{kent2015lorentzian}, including
models of photon dynamics and of photon measurement.    
We represent matter by finitely many massive charged particles,
coupled to the electromagnetic field via interactions that generate 
or absorb finitely many photons.   We suppose the initial state of the 
electromagnetic field also contains finitely many photons, so that,
given the assumption that the photon number changes only finitely, 
the final state also does.   This allows us to describe the
electromagnetic field by photon wave mechanics \cite{
bialynicki1994wave,sipe1995photon,bialynicki1996photon,raymer2005maxwell,muthukrishnan2005photon,smith2006two,
smith2007photon}, which gives a 
physically intuitive picture of the relationship between events
in space-time.    We also comment on possible extensions of these models
into the full quantum field theory regime.    

\section{Inferring local events from late time space-like
  photodetections}

We are interested, initially, in understanding quantum theory in 
Minkowski space-time. 
We now set out some temporary simplifying assumptions,
which should be dropped eventually, 
but which hold in this paper except where explicitly relaxed. 

First, we will neglect gravity.   We take the 
background Minkowski space-time to be fixed.  Our quantum theory does not 
include even a semiclassical treatment of gravity.   
Suppose now that in some finite region $R$ of space-time we have a system
modelled by a finite number $N_m >0$ of massive particles, interacting
via forces that we may model by potentials.   These particles may, for
example, initially be in spatial or energy superpositions.    Suppose also that a finite number $N_p^i \geq 0$ of photons
arrive at this region and interact with the massive particles within
the region.    
We allow that the particles may absorb and/or emit photons, so the
photon number may be variable.   For simplicity, we assume that no
massive particles are created or destroyed, so $N_m$ is constant
throughout.   
The state after the region may contain components with various numbers
of photons leaving the region. 
We will want a significant amplitude (in some examples close to
modulus one) for states with one or more photons leaving the region.
We also assume an upper bound $N_p^o$ on the number of outgoing photons in any
component. 
We allow the possibility $N_p^i = 0$, since we can illustrate
our postulates in simple models in which the state on surfaces prior to the region 
contains no incoming photons.   Realistic models of familiar physical systems
would include both incoming and outgoing photons, of course.   

We suppose further that the above description includes all the matter
(massive particles and photons) in the
space-time, and that all the interactions between photons and massive
particles take place within the region $R$.   We can model this by
taking the Hamiltonian to be local and space-time dependent, with a 
particle-photon interaction that is turned on only 
within $R$.\footnote{Or, as in the models below, we can let the
dynamics effectively ensure that interactions are localized, with
perhaps some exponentially decaying amplitude for late time interactions.} 
Suppose then that we are given an initial state $\ket{\psi_0}$ on some
spacelike hypersurface $S_0$ prior to $R$, which describes $N_m$
massive particles and $N_p^i$ photons.   We suppose we have some
relativistic (albeit spacetime-dependent) unitary
evolution law that allows us to use the Tomonaga-Schwinger formalism. 
We can thus define the evolved state $\ket{\psi_S}$ on any hypersurface
$S$ in the future of $S_0$ via a unitary operator $U_{S_0 S}$. 

\subsection{Intuitive picture}

At an informal intuitive level, then, we have the following picture.  Photons may scatter
from, or be absorbed or emitted by, the massive particles within the 
region $R$.   Scattered and emitted photons then travel away from
$R$, along roughly lightlike paths.   They do not interact further
with massive particles or with one another, so their direction of
propagation outward from $R$ is roughly constant.   On spacelike 
hypersurfaces $S$ in the future of $R$, the outward 
propagating electromagnetic field degrees of freedom will generally be highly
entangled with the massive particle degrees of freedom.  
The outcomes of some measurements of the electromagnetic field will thus
generally be highly correlated with the outcomes of some measurements
of the massive particle states.   

We will work in the centre-of-mass frame coordinates $\ul{x}, t$, 
which we will take for the moment to define a preferred frame. 
Consider now a spacelike hypersurface $S$ comprising three segments,
$S_1 , S_2 , S_3$.   Here $S_1$  runs near to, and in the future of, the future
boundary of $R$, $S_2$ runs close to outward propagating lightlike directions 
from that boundary, and $S_3$ runs outward along the constant time
hyperplane $t=T$, for some large value of $T$.   (In particular,
$T \gg T^R_{\rm max}$, where $ T^R_{\rm max}$ is the largest value of 
the coordinate $t$ of points in $R$.)    
Suppose that we have an infinite array of efficient photodetectors,
each of finite volume $V$, set up along $S_3$.   The volume $V$ is 
chosen to be large enough to allow efficient detection of photons
of frequencies characteristically (i.e. with non-negligible
amplitude) arising in our model, but no larger.   
If we take the detectors to be cubic, we can take the array to 
be regular and to fill $S_3$ (except for a small volume 
around the boundary with $S_2$).  We assume the detectors are 
all stationary in the centre-of-mass frame defining $S_3$, and produce a
measurement result within some characteristic time $\delta$ 
in that frame.     

What should we expect, in this scenario?    The detectors should 
produce a number of clicks, bounded by $N_p^o$, giving an approximate
spatial distribution of detected photons that (may have) interacted 
with the massive particles in the region $R$.   Because both $S$ and
$S_0$ are spacelike hypersurfaces, with $S$ in the future of $S_0$, 
we can calculate the probability distribution for this distribution
from the initial state $\ket{\psi}_{S_0}$ and the unitary evolution 
$U_{S_0 S}$.   Similarly, we can calculate the joint probability 
distribution of possible configurations of clicks and of a
measurement of any operator $A$ defined locally on the surface $S_1$.
In particular, we can do this for operators $A$ defined solely on the
massive particle degrees of freedom, such as the local mass density or 
energy density of these particles.  

\subsubsection{Informal description of the interpretation}

How might we interpret these calculations?   The idea we pursue here is 
that the possible distributions of clicks correspond, roughly
speaking, to ``possible worlds'', or ``possible ways the world 
might turn out'', or ``possible descriptions of reality''.
Each possible distribution of clicks defines an expectation value $\langle A
\rangle$ for the local operator $A$, and these expectation
values, for suitably chosen local operators, give us
variables that augment the quantum state in describing physical
reality on $S_1$.
By applying this procedure for each small region and taking an
appropriate limit, we  obtain variables -- call them beables,
following
Bell -- associated with each point $x$ in space-time.   
These expectation values are functions
of points (or small volumes) in space-time and are to be understood 
as a description of physical reality at these points or in their local
regions.   However, they do not follow local
equations of motion.

To see how this can work, note that, 
while so far we focussed on the physics of region
$R$, and on detectors on the subset $S_3$ of the plane $P_T$ defined
by $t=T$, 
we can imagine similar photodetectors set up throughout $P_T$. 
We consider just a single ``run'' of the ``experiment'' defined
by physics between $S_0$ and $P_T$, and one distribution of clicks
from the detectors distributed over all of $P_T$ defining the
``outcome''.   Then to define the conditional expectation value of a local
operator $A_x$ defined on the massive particle Hilbert space
at point $x$, we condition only on the distribution of clicks from
detectors outside the future light cone of $x$, ignoring any
information (clicks or lack of clicks) about the distribution within
the future light cone.    In this way, by using partial information
from a single full distribution $D$ of clicks, we define a distribution
of local variables $\langle A_x \rangle_D$ throughout space-time, for 
whichever local operators $A$ we choose.   We interpret this as
meaning that each full distribution of clicks $D$ gives us data
corresponding to a ``possible world'', $W_D$, which defines a corresponding
``description of reality'' via the beables  $\langle A_x \rangle_D$. 
Quantum theory, via the Born rule, gives us a probability distribution
$P(D)$ on the full distributions of clicks $D$, and hence on the 
configurations of beables $\langle A_x \rangle_D$ in spacetime.
These distributions depend on the initial state $\ket{\psi}_{S_0}$ and the unitary evolution 
$U_{S_0 S}$, as they must for the picture that emerges to have any
plausible physical meaning.  

To make this concrete, we need to specify the set of operators $A$
from which beables are defined. 
In realistic theories based on QFT, we suggest the stress-energy tensor components $T_{\mu \nu} (x)$ as 
natural candidates.\footnote{The relevant operator here would be the full
  stress-energy tensor, including the stress-energy of the
  electromagnetic field.  Late time detections of photons play a
  special role in our postulates, but this does not require us to
  treat the electromagnetic stress-energy as any ``less real'' than
  that of massive particle fields.   However, in the toy models below,
  we focus on operators describing the states of massive particles, 
  since the models include only one or two photons, which ultimately escape to 
  infinity.}
The charge 4-current components $J_{\mu} (x)$
are another natural possibility.  

The intuition then is that, in realistic models, we expect the beables 
$\langle T_{\mu \nu} (x) \rangle$ to give a description of reality 
that corresponds well with the quasiclassical world we observe.
In particular, they describe high matter densities in the region
occupied by a pointer corresponding to one measurement outcome in a lab
experiment, and very low matter densities in the regions that would
be occupied by pointers corresponding to the other possible outcomes
(assuming these various pointer regions are disjoint).  Moreover,
they do this with the correct Born rule probabilities.   
In Bell experiments, or other experiments on entangled quantum states,
they produced correlated pointer outcomes (in the sense just given)
following standard Bell correlations.    

The reason for this intuition is that the information carried 
away by outgoing photon states, in their position degrees of 
freedom, is correlated with the positions of pointers built
from massive particles with which the photons have interacted. 
So long as large numbers of photons radiate away to future infinity
without reinterfering, this information can be extracted -- using
the conditional expectation value rules given above -- from the 
approximate position measurements made by the photodetector array.     

At this intuitive level, this picture has some
limitations.   It is hard to justify rigorously
in realistic models using QED or any other non-trivial relativistic
quantum field theory, partly because infinite particle number and
photon number states come into play, but also because we have 
no mathematically rigorous definition of non-trivial quantum field
theories and no completely satisfactory way of modelling the dynamics of macroscopic
physical systems and their interaction with microscopic systems within
these theories.    

At first sight it also breaks Lorentz invariance, by 
requiring a frame in which the final photodetector array
is stationary.  For the moment, we take this to be the centre of mass frame for the full unitarily
evolving quantum state (photons plus massive particles). 
Another issue that arises for a full QFT treatment is the choice of photodetector size. 
If we assume that our ideal photodetectors respond only to 
photons of wavelengths somewhat smaller than their size, we 
effectively impose an infra-red cutoff.  
This need not matter in simple models which contain only
finitely many photons and have a lower frequency bound.
However, in realistic models, there is no natural cutoff, 
and the beable configuration probability distribution 
seems likely to depend (at least slightly) on the cutoff choice. 
A further issue is that the beable configuration probability
distribution depends (if only very slightly) on 
the precise configuration of photo-detectors.
 Perhaps this dependence disappears in the limit $T \rightarrow
 \infty$, but that is not clear at this level of discussion. 

\subsection{Implications so far}

It is certainly worth addressing these concerns as far as possible,
and we do consider them further in the rest of the paper.  
However, there is also a case to be made for accepting the picture
at the informal level just given, and developing its scientific
implications.   Two points argue in its favour. 

First, as shown below, in simple but natural models, using versions of
single- or multi-photon quantum theory
derived from quantum field theory, our postulates give 
a well-defined construction that gives probability distributions 
for natural choices of expectation value beables. 

Second, although QED and other relativistic QFTs are not
mathematically
fully rigorously defined, and do not have a complete theory of
measurement, much empirical evidence supports the partial intuitive
understanding of these theories that has been developed.   Calculations fit data with
very high precision; photodetectors can in fact detect single photons
to good approximation.   
Empirical evidence also strongly suggests that our postulates
would extend to work in the observed universe, in the sense that
expectation values for $\langle T_{\mu \nu} (x) \rangle$, fitting
our actual observations, could be reconstructed given an array of 
ideal photodetectors with some lower frequency cutoff on 
a late time hypersurface, stationary with respect to 
cosmological centre-of-mass, using our rules.   That is, 
the expectation value for $\langle T_{\mu \nu} (x) \rangle$
could be reconstructed by conditioning on
photodetector data from outside the causal future of $x$. 

We cannot expect to give a completely rigorous demonstration that this
will work within QFT as presently understood: an absolutely precise and completely rigorous solution of the 
quantum reality problem compatible with relativistic quantum 
field theory will have to await an absolutely precise and completely
rigorous definition of relativistic QFT.  
Even then, to reach our ultimate goal  -- namely, a perfectly and precisely specified realist
version of relativistic quantum theory that also applies to observable
cosmology -- would require a quantum theory of gravity that also (1) 
incorporates a new and rigorous understanding of field theory, (2) comes with a
complete theory of initial conditions {\it and} (3) contains a solution to the quantum
reality problem.  

So, the two points made above may actually go a fair way towards what is
presently achievable in the way of evidence.  (Yet we still hope
to do better!) 

Finally, we would like to underline that  
one of the major aims of this project\cite{kent2012real,kent2014solution,kent2015lorentzian,kent2013beable}
is to use it as a springboard to define testable generalizations \cite{kent1998beyond,kent2013beable} of quantum theory 
that make distinctive experimental and cosmological predictions. 
Since such generalizations may well anyway at best be no more than broadly
correct phenomenological models, we shouldn't necessarily let the finest details of precise
specification hold us back in pursuing them.   A good and well-motivated model that could 
point towards new data is all we need for this purpose, at least.    
It may be most realistic to use well-founded ideas on 
quantum reality to motivate steps forward, without requiring as a
precondition solving every problem in physics at once.  

\section{Photon wave mechanics models}

In a previous paper \cite{kent2015lorentzian}, we illustrated the intuitions underlying our 
proposals with very simple toy models.   In these models, massive particles
were represented by position space wave functions, and photons by
pointlike quantum objects propagating along lightlike trajectories. 
Photon-particle interactions were represented by instantaneous
reflections, which change the direction of photon propagation.
 
Here we give somewhat more realistic models. 
We still assume that there are finite (in our models, small) bounds on
the numbers of photons and massive particles. 
However, we move away from a simplistic picture in which photons
are taken to be pointlike and to follow precisely lightlike paths.   
Instead, we treat them quantum mechanically on an equal footing with 
the massive particles, represented by quantum mechanical wave
functions.   To do this, we use the formalism of photon wave 
mechanics, following in particular results and insights developed by
Bialynicki-Birula \cite{bialynicki1994wave,bialynicki1996photon}, Sipe \cite{sipe1995photon},  Muthukrishnan et al
\cite{muthukrishnan2005photon} and Smith and Raymer
\cite{raymer2005maxwell,smith2006two,smith2007photon}. 
For models with a single photon, we consider
a photon wave function that represents a local probability distribution   
for the photon energy.   
For models with several photons, we consider multi-photon wave functions 
\cite{muthukrishnan2005photon,smith2006two,smith2007photon} which correspond to multi-photon
detection amplitudes and represent states in Fock subspaces of definite
photon number within QED.   
These models allow reasonably realistic descriptions of particle-photon
interactions and of photon emission and absorption by massive
particles.   

We illustrate our proposal by calculating expectation values for
energy state and position beables, which are the natural choices
in the simple models below.  These models have relativistic features -- in
particular, the photon wave functions propagate causally with respect to the Minkowski 
metric -- but are not fully Lorentz invariant.   Fully Lorentz
invariant models would require a more fundamentally satisfactory treatment of 
photon measurements -- a task for the future.  

Examples of photon wave function calculations 
suitable for our purpose were given by Muthukrishnan et al. (MSZ)
\cite{muthukrishnan2005photon}.     MSZ define their photon wave functions in terms of 
the quantized electric field alone.   This allows the wave function
to be directly related to amplitudes for detection in standard
photodetectors, which respond to the electric field.  
Another option \cite{smith2007photon} is to define photon wave functions
in terms of both the quantized electric and magnetic fields.  This 
would be appropriate for (perhaps hypothetical) photodetectors that
are sensitive to both fields.  

We follow MSZ's conventions and closely paraphrase their 
discussion in this section.   MSZ consider the
electric field in a mode volume $V$:  
\begin{eqnarray} 
\hat{E} ( \ul{r}, t ) &=& \hat{E}^+ ( \ul{r} , t ) + \hat{E}^- ( \ul{r}
, t ) \\
&=& \sum_k \sqrt{ ( {\hbar \nu_k }/ {\epsilon_0 V } ) }
( \hat{a}_k u_k ( \ul{r} ) e^{ - i \nu_k t} + \hat{a}_k^{\dagger}
u_k^* ( \ul{r} ) e^{i \nu_k t} ) \, . \nonumber
\end{eqnarray}
Here the $u_k ( \ul{r} )$ are normalised spatial mode functions. 

For a single photon state $\ket{\gamma}$, the probability density of
detecting the photon at $(\ul{r}, t)$ is 
\begin{equation}
\kappa \vert \braopket{ 0 }{ \hat{E}^+ ( \ul{r}, t )}{\gamma} \vert^2 \, , 
\end{equation}
where $\kappa$ is a normalisation constant chosen with dimensions such
that the 
photon wave function 
\begin{equation}
\gamma (\ul{r}, t ) = \sqrt{\kappa} \braopket{ 0 }{ \hat{E}^+ ( \ul{r}, t )}{\gamma}
\end{equation}
has dimension $L^{-3/2}$. 

\subsection{Example 1} MSZ proceed to calculate the wave function $\gamma ( \ul{r} , t )$
describing a photon that is spontaneously emitted by an atom located
at the origin which is introduced at time $t=0$ in an excited state 
and decays at rate $\Gamma$ to its ground state.   They obtain
\begin{equation}\label{gammamode}
\gamma ( \ul{r}, t ) \approx K ( {\sin \theta} / {r} ) \theta ( t - (r/c) )
e^{ -i ( \omega - i \Gamma/2 ) ( t - (r / c ))} \, . 
\end{equation}
Here $c$ is the speed of light, $K$ is a normalisation constant, 
$r = | \ul{r} | $, $\theta$ is the azimuthal angle in coordinates
defined by the axis of the atom's dipole moment, $\omega$ is
the frequency of the atomic transition, and $\theta ( t - (r/c))$ 
is the Heaviside step function, so that $\gamma = 0 $ outside
the future light cone of the spacetime point $(\ul{0}, 0 )$.   
At this point we take the mode volume to tend to infinity, and 
suppose that Eqn. (\ref{gammamode}) applies throughout space
after $t=0$.  

More than one school of thought regarding the interpretation of 
single photon wave functions can be found in the literature.    
It is generally agreed \cite{mandel1966configuration} that
$ \vert \gamma ( \ul{r} , t ) \vert^2 $ is approximately proportional  
to the low rate detection probability of the photon in a suitable photodetector centred
at $( \ul{r} , t )$ with dimensions much larger than the 
typical photon wavelength (given here by $\lambda = c / \omega$).  
Some authors have considered stronger interpretations.
For example, Sipe \cite{sipe1995photon} suggests that ``in attempting to write down a
position-representation wave function, we should be seeking
a probability amplitude $\psi(\ul{r}, t)$ for the photon energy
to be detected about $d\ul{r}$ of $\ul{r}$.''
Bialynicki-Birula \cite{bialynicki1996photon} similarly suggests ``one may introduce a tentative notion of the "average
photon energy in a region of space" and try to associate a probabilistic
interpretation of the photon wave function with this quantity.''
MSZ \cite{muthukrishnan2005photon} ``maintain that a physically meaningful photon wave function
$\gamma( \ul{r}, t)$ can indeed be constructed, that is measurably localized in space, everywhere meaningfully defined
in both phase and amplitude, and provides a valuable tool for understanding photon interference and
correlation experiments.''

We pragmatically adopt the spirit of these suggestions here in order to define 
precise rules, without committing to a final view on the correct 
theory of measurement for photon wave functions or, more generally,
for quantum electrodynamics or relativistic quantum field theories. 
Our discussion is thus meant as an illustrative placeholder, pending a 
more fundamentally satisfactory treatment of 
photon measurements, which we hope to tackle in future work.   

We will adopt as a mathematical hypothesis that an appropriately
normalised $ \vert \gamma ( \ul{r} , t ) \vert^2 $ defines precisely
the detection probability of the photon at $(\ul{r}, t)$, at
asymptotically late times $t$, in the following sense. 
We introduce a hypothetical idealized photodetector 
that is infinitely thin and wide, so that it can occupy a hyperplane. 
We will assume it defines its own preferred frame: we think of it
as composed of idealized particles which have a stationary frame.
In our models, we will take it to occupy some plane $t=T$ ( for
some $T>0$) in the centre-of-mass frame, and assume this coincides with
its own preferred frame.   
We suppose it produces as measurement
output a location $\ul{r}$ at which the photon was detected
at time $T$.
We assume that the photon is absorbed in this process, so 
it has no post-detection state or wave function. 
Moreover, since our real interest is in the asymptotic properties of
the photon wave function at late times, we take the idealized
photodetector to be a construct defined only for the purpose of taking an asymptotic
limit.  This means we need not define any rules at all for physics after it acts.   
We will also assume that the idealized photodetector is 
perfectly efficient.   As the squared amplitude for photon emission tends to
$1$ for large times, this implies that the probability of it 
detecting the photon at some point $\ul{r}$ in space should tend to $1$ as $T \rightarrow
\infty$.  
  
This idealized photodetector is meant as a useful
mathematical fiction.  We do {\it not} claim that it can be
arbitrarily well approximated by real photodetectors, nor that 
it implements standard projective quantum measurements.   
Readers who prefer not to use this construct may revert to 
the standard interpretation of the photon wavefunction as defining
approximate average low rate detection probabilities in suitably large regions, 
and consider detectors set up throughout space at a late time $T$. 
This requires approximations and epsilonics when our postulates
are used to define beables, but otherwise gives a similar picture. 

We have 
\begin{equation}\label{gamma}
\vert \gamma ( \ul{r}, t ) \vert^2 \approx K^2 ( {\sin \theta} / {r} )^2 \theta ( t - (r/c) )
e^{ - \Gamma  ( t - (r / c )) }\, ,
\end{equation}
and our assumptions require 
\begin{equation}
\lim_{t \rightarrow \infty} \int d^3r | \gamma ( \ul{r}, t )
|^2 = 1 \, . 
\end{equation}
We neglect the approximation in the derivation of $\gamma (\ul{r}, t)$, and assume the limit
$L \rightarrow \infty$, taking equation (\ref{gamma}) to be
exact, from now on. 
This gives us
\begin{equation}
K^2 2 \pi = c^{-1} \Gamma \, {\qquad \rm i.e. \, \, } K = \sqrt{ ( {
    \Gamma } / { 2 \pi c}
) } \, . 
\end{equation}

Suppose that the photodetector is in the plane of constant time $T$.   
Our rules imply that the probability of detection at some
position with radial distance 
$ r $ is
\begin{equation}
K^2 ( 2 \pi )  \theta ( T - ( r / c) ) e^{ - \Gamma  ( T - (r / c ))} =
c^{-1} \Gamma \theta ( T - ( r / c) ) e^{ - \Gamma  ( T - (r / c ))} \, . 
\end{equation}
This gives us the probability density for the possible measurement
outcomes at late time $T$.  So the 
probability density for the radial distance of the detection $r$ 
satisfying $ r =  c (T -  t)$ is $c^{-1} \Gamma \theta(t) e^{-\Gamma t
}$, and the probability of $r$ lying in the range $\lbrack cT - c(t+dt), cT
- ct \rbrack$ is $\Gamma \theta(t) e^{-\Gamma t} {dt} $.   

We ignore recoil and spreading of the atom position space wave
function, taking an idealised model in which the atom remains fixed at $\ul{r}=\ul{0}$ for all $t$. 
The quantum state of the atom plus photon at time $t > 0$ is then given by
\begin{equation}
\ket{e} e^{- \Gamma t /2}  + \ket{g} \gamma( \ul{r}, t ) \, ,
\end{equation}
where $\ket{e}$ is the excited state of the atom and $\ket{g}$ the
ground state.   Consider now the hypersurface defined by the light cone
from $(\ul{0}, t)$ to late time $T$, joined with the outward part
of the time $T$ hyperplane.   The effective quantum state on this
hypersurface, defined as the limit of the quantum states on spacelike
hypersurfaces that tend to it, is 
\begin{equation}
\ket{e} e^{ - \Gamma t /2 } + \ket{g} \gamma ( \ul{r}, T ) \theta ( r
- (T - t ) c ) \, . 
\end{equation} 
In summary, a detection on the final hyperplane at radius $r = cT - c
t_0$ occurs with probability density  $\Gamma \theta(t_0 ) e^{-\Gamma
  t_0 }$.   The detection takes place on the relevant limiting
hypersurface for defining the beables if $t \geq t_0$. 
Our postulates imply that, given such a detection, the atom energy
state beable is $\ket{e} \bra{e}$ for $ t < t_0$ and $\ket{g} \bra{g}$
for $t \geq t_0$.      In other words, the atom energy state beable
undergoes a transition from excited state to ground state at a
definite time $t_0$, which is randomly selected, with 
probability density  $\Gamma \theta(t_0 ) e^{-\Gamma   t_0 }$. 

Of course, from the perspective of observers within the system, who
are unaware of the outcome of the final measurement (both because
it occurs at late times and because it is anyway fictional!), the 
beable transition time is not predictable.    
Such an observer can only assign the beable the 
mixed state
\begin{equation}   ( 1  - e^{ - \Gamma t_0 }) \ket{g} \bra{g} + e^{ - \Gamma t_0 } \ket{e}
\bra{e} \, . \end{equation}
Nonetheless, on the view we are advocating, this subjective 
mixed state describes the observer's incomplete knowledge of 
an objective fact.   At any given time $t_0$, the atom energy
beable is either in the state $ \ket{e} \bra{e}$, or has transitioned
to state $\ket{g} \bra{g}$.  In the latter case, in this simple model,
it remains in this
state at all later times.    

\subsection{Example 2}

Suppose now that at time $t=0$ the atom is in a superposition of 
two position states, which we call $ \ket{\pm d/2}$.  
These have centre of mass at $\ul{r}_{\pm} = (0, 0, \pm d/2 )$, where
the atomic dipole is aligned with the $z$ axis.      
As before, it is introduced at $t=0$ in the excited state $\ket{e}$.  
We thus take the initial atom state to be 
\begin{equation}
\alpha \ket{ - d/2} \ket{e} + \beta \ket{  d/ 2} \ket{e} \, . 
\end{equation} 
As before, we will ignore recoil and the spread of the atom
position space wavefunction.
Thus the time $t$ atom-photon state is 
\begin{equation}
\alpha ( \ket{ - d/2} \ket{e} e^{-\Gamma t / 2} + \ket{-d/2} \ket{g}
\gamma ( \ul{r} - \ul{r}_- , t) )  + 
\beta  ( \ket{ d/ 2} \ket{e} e^{- \Gamma t / 2} + \ket{d/2} \ket{g} ) \gamma
(\ul{r} - \ul{r}_+ , t )  \, . 
\end{equation}
We take $  d \gg \lambda $, where as above $ \lambda = c /
\omega$ is the characteristic wavelength, and 
  $ \Gamma \ll \omega$. 
For   $ \Gamma \ll \omega$, the overlap is \footnote{This follows from
MSZ's given calculation method and agrees with their plot in their
Figure 2(b).   In particular, it 
gives overlap $1$ at $d=0$, as expected.  MSZ's stated result, 
given in their equation (14), appears to contain typos.}   
\begin{equation}
\int d \ul{r} \gamma^* ( \ul{r} - \ul{r}_- , t) )  \gamma
(\ul{r} - \ul{r}_+ , t )  =  { { 3 ( \sin ( 2 \pi d / \lambda ) - ( 2 \pi d /
  \lambda ) \cos ( 2 \pi d / \lambda ))  } \over {  ( 2 \pi d / \lambda
  )^3 }} \, , 
\end{equation}
which is small for $d \gg \lambda$. 

At large time $T$, our idealized photo-detector will register a 
photon either in the effective support of $\gamma ( \ul{r} - \ul{r}_-
, T)$ or in that of $\gamma ( \ul{r} - \ul{r}_+ , T)$, and these
supports are almost disjoint.   
Neglecting small contributions from
detections at points where $ \vert \gamma ( \ul{r} - \ul{r}_+ , T) \vert^2
\approx \vert \gamma ( \ul{r} - \ul{r}_- , T) \vert^2$, we find a detection
in the effective support of the former with probability $ \vert \alpha
\vert^2$, and the latter with probability $ \vert \beta \vert^2$.

In the first case, a detection at a point $\ul{r}$, with 
$ \vert \ul{r} - \ul{r}_+ \vert = c (T - t)$, implies that the 
energy beable at $ ( \ul{r}_+ , t )$ transitions from
$ \ket{e} \bra{e}$ to $\ket{g} \bra{g}$, as
above.  However, the position beable also transitions at that point, 
so that the total local contribution to the energy at $\ul{r}_+$ transitions from
$ \vert \alpha \vert^2 \ket{e} \bra{e}$ to $\ket{g} \bra{g}$. 
At $\ul{r}_-$, the transition from $ \vert \beta \vert^2 \ket{e} \bra{e} $ 
to (approximately) a zero density matrix takes place at $t'$, where 
$( \ul{r}_- , t' )$ is lightlike separated from $ ( \ul{r}, T) $. 
This gives us 
\begin{equation}
t' = t - { 1 \over c} \hat{\ul{n}}{\bf \cdot } ( \ul{r}_+ - \ul{r}_- )
+ O ( { 1 \over T} ) \, ,
\end{equation}
where $ \hat{\ul{n}}$ is the unit vector in the spatial direction
separating the detection point from  $( \ul{r}_+ , t )$.
So, in the asymptotic limit $T \rightarrow \infty$, $t'$ depends only on 
$t$ and on $ \hat{\ul{n}}$. 

Similarly, if the detection is in the effective support of 
$\gamma ( \ul{r} - \ul{r}_- , T)$, the energy and position beables
at $\ul{r}_-$ transition to approximately $\ket{g} \bra{g}$, and those at
$\ul{r}_+$ to approximately zero.   

In other words, the beables describe an effective transition from
position superposition to (approximately) one component of the
superposition.   This takes place in a frame that depends on the
photo-detection result in a way that has a well-defined asymptotic limit.  

\subsection{Example 3}

Again following MSZ, consider a cascade decay of an atom introduced 
at the origin at $t=0$ in a level two excited
state, which decays via a level one excited state to the ground state, emitting
two photons with characteristic frequencies $\omega_1$ and $\omega_2$
at rates $\Gamma_1$ and $\Gamma_2$.  
The two-photon wave function \cite{muthukrishnan2005photon,smith2006two,smith2007photon} can be defined
from the field state $\ket{\Psi}$ by 
\begin{equation}
\Psi ( \ul{r_1} , t_1 ; \ul{r_2} , t_2 ) = \sqrt{\kappa'} \braopket{0}{ \hat{E}^+
(\ul{r_2}, t_2 )    \hat{E}^+ (\ul{r_1}, t_1 ) }{ \Psi } \, . 
\end{equation}
We suppose that, in a photon number eigenstate with $N_{\rm \gamma} =2$, our idealized
photodetector lying in the hyperplane $t=T$ will always detect two
photons, producing as measurement outputs the pair of positions    
$\ul{r_1}$ and $\ul{r_2 }$ with probability 
$ \vert \Psi ( \ul{r_1 } , T ; \ul{r_2} , T ) \vert^2 $, 
which gives us a normalisation condition by integrating over
$\ul{r_1}$ and $\ul{r_2}$.
Again, we stress that our measurement rules for the idealized
photodetector are mathematical fictions, not arbitrarily well
approximated by real photodetectors.  

MSZ obtain
\begin{eqnarray}
\psi ( \ul{r_1} , t_1 ; \ul{r_2} , t_2 ) &=& K' ( { \sin \theta_1 } / r_1 ) (
{\sin \theta_2 } / r_2 ) \theta ( t_1 - ( r_1 / c )) e^{- ( i \omega_1
  + ( \Gamma_1 / 2) ) ( t_1 - (r_1 / c))}   \\
&& \theta (( t_2 - (r_2 / c)) - (t_1 - (r_1 / c ))) e^{- ( i \omega_2
  + ( \Gamma_2 / 2))( (t_2 - (r_2 / c )) - (t_1 - (r_1 / c)))}
\nonumber \\
&&+ ( 1 \leftrightarrow 2 ) \, . \nonumber 
\end{eqnarray}

Taking $t_1 = t_2 =  T$ allows us to take $r_2 < r_1$ without loss of
generality.   We obtain, as above, the probability of detection of at least one 
photon at time $T$ and radial distance $r_1  > (T-t)c$ to be 
$( 1  - e^{ - \Gamma_1 t}) $, for $ 0 < t < T$.   
Conditioned on detecting at least one photon, and with $r_1$ as
the maximum radial distance of detection, the
probability of detection of a second photon at radial distance
$r_2 < r_1$ is $( 1 - e^{ - \Gamma_2 ((r_2 - r_1)/ c )} )$. 
Again, these results are independent of $T$, so the asympotic limit
for $T \rightarrow \infty$ is well defined. 
According to our postulates, the atom energy beables now transition
from the second excited state to the first excited state, and then from the first
excited state  to the ground state.  These transitions almost
certainly occur by time $t$ within a few multiples of $ (
\Gamma_1^{-1} + \Gamma_2^{-1} )$. 

\subsection{Example 4}

Consider again the atom of Example 1, introduced in an excited state 
at $( \ul{r}, t) = ( \ul{0} ,  0 )$.     Suppose now that we have 
a larger massive object, which acts as an essentially perfect absorber
of photons over a range of frequencies
including the atom's characteristic transition frequency $\omega$.
Suppose that the object is spherical, with radius $R \gg 10
\Gamma^{-1} c$, and suppose it is initially in a superposition
of two stationary states, which we call $\ket{0}_{\rm obj}$ and
$\ket{100}_{\rm obj}$, with centres of mass $\ul{r}_0 = \ul{0}$ 
and $\ul{r}_{100} = ( 100 R , 0 , 0 )$ respectively.   As with the 
atom, we neglect the spread or evolution of the object's 
position space wave function, and we also neglect any momentum
transfer from photon absorption.       We also assume the atom
can coexist with the object in state $\ket{0}_{\rm obj}$ without any 
interaction other than the emission and subsequent absorption
of a single photon, although they have the same centre-of-mass.

To make this more plausible, we may take the object to be 
a thick uniform spherical shell with outer radius $R$ and inner radius
$R' < \Gamma^{-1} c$.   To make it even more plausible, we may allow
a small hole in the shell, so that the atom can be introduced 
at $( \ul{0} , 0)$ without interacting with the $\ket{0}_{\rm obj}$ component 
of the object, and thus without initially affecting the superposition.
We neglect the small amplitude of the emitted photon propagating
through the hole in the calculations below. 

The initial combined state is thus
\begin{equation}
\ket{e}_{\rm atom} ( \alpha \ket{0}_{\rm obj} + \beta \ket{100}_{\rm
  obj} ) \, .
\end{equation}
At late time $T$, this evolves to approximately  
\begin{equation}
\ket{g}_{\rm atom} ( \alpha \ket{0^*}_{\rm obj} + \beta \ket{100}_{\rm
  obj} \gamma (\ul{r} , T ) \, 
\end{equation}
where $\ket{0^*}$ is the state of the object at centre of mass $\ul{r}
= \ul{0}$ after having absorbed a photon, $\gamma ( \ul{r}, T )$ is 
the photon wave function calculated as in Example 1, and we neglect
small components arising from the object in state $\ket{100}_{\rm obj}$
absorbing a photon propagating close to the positive $x$ direction, 
and also neglect the exponentially small amplitude of the atom remaining in excited state 
at time $T$. 

Our postulates now imply that at late time $T$ the beable describing
the atom energy has evolved to $\ket{g} \bra{g}$, as before. 
Our ideal photo-detector will detect a photon with probability
$ \vert \beta \vert^2$, in which case our postulates imply beable 
$\ket{100}_{\rm obj} \bra{100}_{\rm ojb}$ for the object.   Alternatively, with
probability $ \vert \alpha \vert^2$, there is no photo-detection, 
and our postulates imply beable $\ket{0^*}_{\rm obj} \bra{0^*}_{\rm obj}$ for the object.

In other words, our postulates describe a transition for the object
beables, transitioning to one of the superposition components, with 
the standard Born rule probabilities.   

\section{QFT and Infrared cutoffs}

Our models so far have been within a version of quantum theory in
which the photon propagation respects the Minkowski causal structure,
but which relies on approximations and which is valid only in a 
restricted regime.   Modelling within full relativistic quantum
field theory poses several problems.   Among these is that
realistic physical processes involving 
charged particles interacting with an electromagnetic field 
generally generate an indefinite number of low energy photons.    
Including these in a model directly would
mean we could not restrict to states of finite photon number, 
nor assume that our late time idealized photodetector would 
produce only finitely many photodetections.   

Standard treatments of QED solve the infrared divergence problem 
by noting that real world detectors cannot detect photons below
some cutoff energy.   The amplitudes corresponding to photon
emissions below this energy can thus be combined with amplitudes
involving low energy virtual photons, producing a finite sum
from two contributions that would separately be infinite. 
This suggests possible ways to treat the infrared divergence 
in our model.

Perhaps the simplest option is to impose an infrared cutoff in
the calculations for each late time $T$ hypersurface, so that
photon states of frequency less than $ \nu (T)$ are neglected, with
$\nu(T) \rightarrow 0$ as $T \rightarrow \infty$. 
This would imply a finite number of photodetections on any
given late time hypersurface, although that number would generally
be expected to tend to infinity as the hypersurface moves towards
future infinity.    Taking photodetections of photons with 
average frequency $\nu (T)$ to give an outcome with position uncertainty
at least some multiples of the wavelength $\lambda(T) = c / \nu(T)$
suggests requiring $\nu(T) \approx N c/ T$, for not very large $N$, in the relevant
frame.
Intuition then suggests that (i) soft photon emissions, 
from bremstrahlung or other processes, should correlate with the 
states of matter in a very similar way to the emissions and 
scatterings of hard photons, modelled above, 
(ii) the beable expectation values inferred from detections of soft
and hard photons ought to be approximately consistent, in a way
which tends to a stable limit as $T \rightarrow \infty$ and 
$\nu(T) \rightarrow 0$.  
Testing this intuition in models that include bremstrahlung is a task
for the future.  

As an approximation, one could replace our idealized
photodetector at late time $T$ with a more realistic (though of course, still fictional) array
of adjacent photodetectors, which have finite volumes $V \approx L(T)^3$, have a
centre-of-mass frame, and which only detect photons above a certain
cutoff frequency $\nu (T)$ with respect to that frame.
This requires $L(T) = N  \lambda (T)$ for some large, but not
necessarily very large, $N$.  
We would expect that, even if $\nu(T) \rightarrow \nu_{\rm min} > 0$, 
the beables should define quasiclassical physics in
realistic models, if the cutoff $\nu_{\rm min}$ is small enough. 

This picture has the advantage that it
is empirically supported. 
The behaviour of real photodetectors is well understood, and we are pretty confident 
that, all else being equal, we can sensibly speak of the sort of
readings we would expect from an array of them at late times, either 
in Minkowski space models or in the universe we observe.
However, it seems hard to make it perfectly precise.
Since realistic photodetectors produce an approximate measurement output
corresponding to an unsharp measurement of the photon position, this
 introduces a corresponding uncertainty in the definition of the 
beable expectation values.     
Imposing a finite infrared cutoff also appears to break
Lorentz invariance, and the photodetectors' volumes and other
characteristics require further arbitrary parameter choices.   Although
the picture of quasiclassical physics defined by the beables need not
be very sensitive to the precise choices made, within sensible
ranges, the precise beable values nonetheless will depend at least
slightly on these choices.  

An underlying issue here is that
while in principle relativistic quantum field theory is a Lorentz
covariant dynamical theory, and while the recipes that are used to
extract experimental predictions from it are consistent with Lorentz
symmetry, we do not presently have a well-defined Lorentz covariant
theory of approximately localized measurements in quantum field theory.  
Our resorting here to idealizations or approximations in defining photodetectors reflects
this.   Some of the well-known no-go theorems 
\cite{reeh1961bemerkungen,malament1996defense,halvorson2002no} may
perhaps be specific to algebraic quantum field theory.   
However the lack of rigorous treatments of approximately localized
measurements within other formulations of QFT suggests that every 
formulation of QFT currently has significant gaps
in its conceptual framework. 

As already noted, for the purpose of defining testable generalizations of quantum theory 
that make distinctive experimental and cosmological predictions, but
which at best are only roughly correct, a lack of perfect precision
may not matter.  
And perhaps complete precision is not possible until we get beyond the
current formulation of quantum field theory and/or improve the
associated theory of localized measurements. 
That said, of course an absolutely precisely specified theory,
without any arbitrary parameter choices, would be preferable if available. 

\section{Final time photon frequency measurements} 

Another possible solution to the infrared problem is to drop 
the specific proposal of Ref \cite{kent2015lorentzian} and use instead
ideas outlined in \cite{kent2012real,kent2014solution}, and consider 
late time measurements of the momentum distribution of photons rather than approximate or
exact position measurements.   

The photon wave functions and number operators are 
well-defined and local in momentum space.  We can thus simply postulate
that a standard (idealised) quantum measurement takes place on a 
late time hyperplane $t=T$ in some frame, producing as outcome
a set of photon momenta $\ul{p}_1 , \ul{p}_2 , \ldots$.   This set may
be infinite, and the $\ul{p}_i$ may not necessarily be distinct.  
        
Instead of defining beables at a point $x$ via the construction 
of Ref. \cite{kent2015lorentzian}, described above, we use the ABL
rule \cite{aharonov1964time} 
construction of Refs. \cite{kent2012real,kent2014solution}.   The 
beable expectation values at a point $x$ are defined on 
hypersurfaces asymptotically tending to the past surface of $x$.
To do this, we use the initial state on the initial hypersurface
as pre-selected initial data, and the outcome of the final momentum measurement
on the late time hyperplane as post-selected final data.   
Here, as above, we use the final hypersurface photon data to define expectation value
beables for the massive particle degrees of freedom.    

The intuition is that this renders the infrared problem
innocuous, since (i) the ABL rule allows us to make inferences about the massive
particle expectation value beables 
from the distribution of the photons above a lower cut-off frequency,
(ii) these inferences are either essentially confirmed, or else unaffected, by
including the distribution of soft photons in the final measurement outcome.  
That is, either the distribution of soft photons is essentially
uncorrelated with the massive particle states, in which case including 
that distribution makes essentially no difference, or it is usefully correlated.
Either way, in realistic models, we expect the beables 
inferred from photons above a low energy cut-off to already
characterise the expectation value beables fairly precisely, and
including the entire photon spectrum to make only small corrections.    

As with our earlier examples, testing this intuition in models that include bremstrahlung is a task
for the future.  However, putting infrared problems aside, we can
illustrate the basic proposal with another toy example: 

\subsection{Example 5}

Consider again the atom-object system of Example 4.
The initial combined state is
\begin{equation}
\ket{e}_{\rm atom} ( \alpha \ket{0}_{\rm obj} + \beta \ket{100}_{\rm
  obj} ) \, .
\end{equation}
At late time $T$, this evolves to approximately  
\begin{equation}
\ket{g}_{\rm atom} ( \alpha \ket{0^*}_{\rm obj} + \beta \ket{100}_{\rm
  obj} \gamma (\ul{r} , T ) \, 
\end{equation}
if we represent the photon by a position space wave function.
Representing the photon by a momentum space wave function instead,
we have 
\begin{equation}
\ket{g}_{\rm atom} ( \alpha \ket{0^*}_{\rm obj} + \beta \ket{100}_{\rm
  obj} \gamma (\ul{p}, T ) \, 
\end{equation}
where $\gamma (\ul{p}, T )$ is sharply peaked around momenta
for which $\vert \ul{p} \vert  = h \nu / c $.  
A late time hyperplane measurement of the photon momentum
distribution, in this model, will either (with probability 
$\vert \alpha \vert^2$ ) produce no photon, or (with probability
$\vert \beta \vert^2 $) produce one photon with momentum $\ul{p}$, 
obtained randomly from the distribution  $ \vert \gamma ( \ul{p} , T )
\vert^2$.   From the ABL rule, we infer object position expectation value
beables $\ket{0^*} \bra{0^*}$ and
$\ket{100} \bra{100}$ respectively, at any time when the object
is in the quantum superposition above.   (We also infer 
atom energy expectation value beables $\ket{g}\bra{g} (1 - e^{-\Gamma t} ) + \ket{e} \bra{e}
e^{-\Gamma t}$, regardless of the photon momentum measurement
outcome.)   

In realistic models, 
roughly localised collections of photons with different momentum
distributions will be found in different regions.   In our universe,
this is so because of the uneven distribution of galaxies, stars, and
inhomogeneities arising at early times after the Big Bang; on Earth, 
of course,  the distribution of smaller energy sources is also uneven.  
This model illustrates that these distributions, combined with a 
late time photon momentum distribution measurement, can allow 
a quasiclassical description of massive particles with which
the photons (may) interact.   Note that we would get the
same qualitative result if we allowed the object to re-emit
the photon energy after absorption, so long as it emitted
a different spectrum (for example, several photons of lower
energy).    

A striking feature of this example, in comparison to example $4$, 
is that the object expectation value beables are assigned
to one or other component of the position superposition 
as soon as that superposition is created, even if the 
interaction with the atom occurs much later.    
In examples $1$-$4$, we used a construction \cite{kent2015lorentzian} that
makes inferences about the beables at $x$ only
from final time measurement results outside
the future light cone of $x$.  
In this example, following the ideas of
Refs. \cite{kent2012real,kent2014solution},
 we use measurement outcomes on the electromagnetic field from 
the entire late time hypersurface to define beables for the
massive particles, via the ABL rule \cite{aharonov1964time}.   
In some sense, in this version, real (not just fictional late time) 
future measurement interactions between photons and massive
particles cause past measurement outcomes for the massive particles.
This feature may be counterintuitive or unaesthetic to some, but it is 
not clear that {\it per se} it is fundamentally problematic. 

\section{Conclusions} 

It has long been well understood that photons very effectively 
decohere quantum systems (e.g. \cite{joos1985emergence,joos2013decoherence,riedel2010quantum}). 
Our simple models could be made more realistic by drawing on  
these and other earlier analyses.  
The point of our models, however, is not to give another illustration
of decoherence, but rather to illustrate that a coherent picture of quantum reality
can be constructed from a single postulate \cite{kent2015lorentzian}.
This relies on the observation 
that a significant fraction of photons scattered from massive 
objects will propagate away to future infinity, with little or 
no further scattering.   As our models illustrate, given the 
postulate and the right type of initial conditions, a  
single photon scattering to infinity can suffice.   
In our cosmology, we expect, the generic anisotropy of the photon environment 
supplies the right type of initial conditions and the relatively sparse
distribution of matter allows some fraction of photons to propagate
with little or no scattering.  The apparently indefinite future
expansion, with the corollary that interactions become less and less
frequent, appears to allow the right asymptotic behaviour for the final time 
measurement postulate to be well defined.\footnote{It should be
acknowledged that there is theoretical and empirical uncertainty about the 
true asymptotic final cosmological state.   For example, if, as some imagine, it 
reaches a true vacuum state, quasiclassical physics in earlier eras
could not be reconstructed from our asymptotically late time
measurement postulate.}  

We have thus illustrated the proposal of Ref. \cite{kent2015lorentzian} in
some simple but (in their bare essence) qualitatively realistic quantum mechanical models, and
shown it works satisfactorily in these models, giving 
natural realist beable picture for finite time physics from 
hypothetical asymptotically late time measurements on the 
electromagnetic field.    We discussed the problem of infra-red
effects, and proposed possible solutions.     
For comparison, we have also illustrated an alternative proposal 
\cite{kent2012real,kent2014solution}, which offers another possible
way of addressing infrared problems and gives a qualitatively
different, though also apparently consistent, beable picture.      

These illustrations progress well beyond the toy models
of Ref. \cite{kent2015lorentzian}.
More detailed models, and in particular models incorporating a  
a more fundamentally satisfactory treatment of photon measurements,
would still be desirable in developing this work further. 

Even at the present level of development, though, we have 
set out a simple proposal for a rule that 
allows branches of the universal wave function to be 
approximately characterised by fictitious late time measurements of the electromagnetic
field in realistically modelled photodectors.   
Again we stress that, while this would not work in every cosmology, it should in 
an indefinitely expanding universe like ours in which interactions
become less and less frequent at late times. 

This gives us a natural one-world alternative to 
Everettian quantum theory, according to which only one
branch is realised, chosen randomly via the Born rule 
applied to these late time measurements. 
The branching structure in our approach is most naturally described in terms of 
evolving expectation value beables, rather than  
(for example) sequences of orthogonal projective measurements.
In our opinion, it compares well to other approaches, offering a much
simpler description of quasiclassical physics than achieved (
at least to date) by any set selection rule in the consistent
or decoherent histories approach to quantum theory
\cite{gell1990quantum,gell1993classical,dowker1996consistent,gell2012decoherent} for example. 

The models presented here have supposed that beables throughout space-time can be
constructed from fictitious asymptotic late time measurements of
photons.   This is a natural choice when considering physics in
Minkowski space or other fixed background space-times, assuming that
photons are indeed massless, since if photons propagate outside the future
light-cone $L_x$ of a point $x$ they do not return, 
absent rescattering that reflects them back into $L_x$.   Assigning this role to gravitons might be an 
even more natural choice in a quantum theory of gravity.
There are at least two reasons for this. 
First, postulating a fundamentally different role for matter and
gravitational degrees of freedom is arguably more natural than one
between massive and massless particles.
Second, rescattering and reflection back of gravitons 
into the future light cone is less generic. 
In particular, good photon mirrors
exist, but gravitational mirrors (in the standard sense)
do not.  

However, the nonlinearities of quantum gravity make it hard to set out  
a conceptually consistent discussion, since the future causal
structure itself is affected by radiated gravitons. 
Without a full quantum gravity theory, it is in any case hard to give
convincing illustrations in simple models.   Nonetheless, it would be
interesting to explore this possibility further.

\section{Acknowledgements}
I gratefully acknowledge the support of a Foundational Questions
Institute  (FQXi) grant. 
This work was partially supported  by Perimeter Institute
for Theoretical Physics. Research at Perimeter Institute is supported
by the Government of Canada through Industry Canada and by the
Province of Ontario through the Ministry of Research and Innovation.
I thank Michael Raymer and John Sipe for helpful discussions.  

\section*{References}

\bibliographystyle{unsrtnat}
\bibliography{photonreality}{}

\begin{thebibliography}{24}
\providecommand{\natexlab}[1]{#1}
\providecommand{\url}[1]{\texttt{#1}}
\expandafter\ifx\csname urlstyle\endcsname\relax
  \providecommand{\doi}[1]{doi: #1}\else
  \providecommand{\doi}{doi: \begingroup \urlstyle{rm}\Url}\fi

\bibitem[Kent(2012)]{kent2012real}
Adrian Kent.
\newblock Real world interpretations of quantum theory.
\newblock \emph{Foundations of physics}, 42\penalty0 (3):\penalty0 421--435,
  2012.

\bibitem[Kent(2014)]{kent2014solution}
Adrian Kent.
\newblock Solution to the {L}orentzian quantum reality problem.
\newblock \emph{Physical Review A}, 90\penalty0 (1):\penalty0 012107, 2014.

\bibitem[Kent(2015)]{kent2015lorentzian}
Adrian Kent.
\newblock Lorentzian quantum reality: postulates and toy models.
\newblock \emph{Phil. Trans. R. Soc. A}, 373\penalty0 (2047):\penalty0
  20140241, 2015.

\bibitem[Bialynicki-Birula(1994)]{bialynicki1994wave}
Iwo Bialynicki-Birula.
\newblock On the wave function of the photon.
\newblock \emph{Acta Physica Polonica-Series A General Physics}, 86\penalty0
  (1):\penalty0 97--116, 1994.

\bibitem[Sipe(1995)]{sipe1995photon}
JE~Sipe.
\newblock Photon wave functions.
\newblock \emph{Physical Review A}, 52\penalty0 (3):\penalty0 1875, 1995.

\bibitem[Bialynicki-Birula(1996)]{bialynicki1996photon}
Iwo Bialynicki-Birula.
\newblock The photon wave function.
\newblock In \emph{Coherence and Quantum Optics VII}, pages 313--322. Springer,
  New York, 1996.

\bibitem[Raymer and Smith(2005)]{raymer2005maxwell}
MG~Raymer and Brian~J Smith.
\newblock The {M}axwell wave function of the photon.
\newblock In \emph{Optics \& Photonics 2005}, pages 293--297. International
  Society for Optics and Photonics, 2005.

\bibitem[Muthukrishnan et~al.(2005)Muthukrishnan, Scully, and
  Zubairy]{muthukrishnan2005photon}
A~Muthukrishnan, MO~Scully, and MS~Zubairy.
\newblock The photon wave function.
\newblock In \emph{Optics \& Photonics 2005}, pages 287--292. International
  Society for Optics and Photonics, 2005.

\bibitem[Smith and Raymer(2006)]{smith2006two}
Brian~J Smith and MG~Raymer.
\newblock Two-photon wave mechanics.
\newblock \emph{Physical Review A}, 74\penalty0 (6):\penalty0 062104, 2006.

\bibitem[Smith and Raymer(2007)]{smith2007photon}
Brian~J Smith and MG~Raymer.
\newblock Photon wave functions, wave-packet quantization of light, and
  coherence theory.
\newblock \emph{New Journal of Physics}, 9\penalty0 (11):\penalty0 414, 2007.

\bibitem[Kent(2013)]{kent2013beable}
Adrian Kent.
\newblock Beable-guided quantum theories: Generalizing quantum probability
  laws.
\newblock \emph{Physical Review A}, 87\penalty0 (2):\penalty0 022105, 2013.

\bibitem[Kent(1998)]{kent1998beyond}
Adrian Kent.
\newblock Beyond boundary conditions: General cosmological theories.
\newblock In L.~Roszkowski, editor, \emph{Particle Physics and the Early
  Universe, Proceedings of COSMO-97}, pages 562--564, New Jersey, 1998. World
  Scientific.

\bibitem[Mandel(1966)]{mandel1966configuration}
L~Mandel.
\newblock Configuration-space photon number operators in quantum optics.
\newblock \emph{Physical Review}, 144\penalty0 (4):\penalty0 1071, 1966.

\bibitem[Reeh and Schlieder(1961)]{reeh1961bemerkungen}
Helmut Reeh and Siegfried Schlieder.
\newblock Bemerkungen zur unit{\"a}r{\"a}quivalenz von lorentzinvarianten
  feldern.
\newblock \emph{Il Nuovo Cimento (1955-1965)}, 22\penalty0 (5):\penalty0
  1051--1068, 1961.

\bibitem[Malament(1996)]{malament1996defense}
David~B Malament.
\newblock In defense of dogma: Why there cannot be a relativistic quantum
  mechanics of (localizable) particles.
\newblock In \emph{Perspectives on quantum reality}, pages 1--10. Springer,
  1996.

\bibitem[Halvorson and Clifton(2002)]{halvorson2002no}
Hans Halvorson and Rob Clifton.
\newblock No place for particles in relativistic quantum theories?
\newblock \emph{Philosophy of Science}, 69\penalty0 (1):\penalty0 1--28, 2002.

\bibitem[Aharonov et~al.(1964)Aharonov, Bergmann, and
  Lebowitz]{aharonov1964time}
Yakir Aharonov, Peter~G Bergmann, and Joel~L Lebowitz.
\newblock Time symmetry in the quantum process of measurement.
\newblock \emph{Physical Review}, 134\penalty0 (6B):\penalty0 B1410, 1964.

\bibitem[Joos and Zeh(1985)]{joos1985emergence}
Eric Joos and H~Dieter Zeh.
\newblock The emergence of classical properties through interaction with the
  environment.
\newblock \emph{Zeitschrift f{\"u}r Physik B Condensed Matter}, 59\penalty0
  (2):\penalty0 223--243, 1985.

\bibitem[Joos et~al.(2013)Joos, Zeh, Kiefer, Giulini, Kupsch, and
  Stamatescu]{joos2013decoherence}
Erich Joos, H~Dieter Zeh, Claus Kiefer, Domenico~JW Giulini, Joachim Kupsch,
  and Ion-Olimpiu Stamatescu.
\newblock \emph{Decoherence and the appearance of a classical world in quantum
  theory}.
\newblock Springer Science \& Business Media, 2013.

\bibitem[Riedel and Zurek(2010)]{riedel2010quantum}
C~Jess Riedel and Wojciech~H Zurek.
\newblock Quantum darwinism in an everyday environment: Huge redundancy in
  scattered photons.
\newblock \emph{Physical review letters}, 105\penalty0 (2):\penalty0 020404,
  2010.

\bibitem[Gell-Mann and Hartle(1990)]{gell1990quantum}
Murray Gell-Mann and James~B Hartle.
\newblock Quantum mechanics in the light of quantum cosmology.
\newblock In W.~Zurek, editor, \emph{Proceedings of the {S}anta {F}e
  {I}nstitute {W}orkshop on {C}omplexity, entropy and the physics of
  information}, volume~8, Redwood City, 1990. Addison Wesley.

\bibitem[Gell-Mann and Hartle(1993)]{gell1993classical}
Murray Gell-Mann and James~B Hartle.
\newblock Classical equations for quantum systems.
\newblock \emph{Physical Review D}, 47\penalty0 (8):\penalty0 3345, 1993.

\bibitem[Dowker and Kent(1996)]{dowker1996consistent}
Fay Dowker and Adrian Kent.
\newblock On the consistent histories approach to quantum mechanics.
\newblock \emph{Journal of Statistical Physics}, 82\penalty0 (5-6):\penalty0
  1575--1646, 1996.

\bibitem[Gell-Mann and Hartle(2012)]{gell2012decoherent}
Murray Gell-Mann and James~B Hartle.
\newblock Decoherent histories quantum mechanics with one real fine-grained
  history.
\newblock \emph{Physical Review A}, 85\penalty0 (6):\penalty0 062120, 2012.

\end{thebibliography}
\end{document}